\documentclass[a4paper]{article}

\usepackage{amssymb}
\usepackage[nodots]{numcompress}
\usepackage{amsmath,amstext,amsgen,amsbsy,amsopn,amsfonts,amssymb}
\usepackage{picins}
\usepackage{amsthm}
\usepackage{bm}
\newtheorem{theorem}{Theorem}
\newtheorem{lemma}{Lemma}

\usepackage{graphicx} 
\usepackage{subfigure}
\usepackage{float}
\usepackage{algorithm}
\usepackage{algorithmic}

\begin{document}

\title{\textbf{Cluster Forests}\footnotetext{Author contact: dhyan@berkeley.edu, Aiyouchen@google.com, jordan@cs.berkeley.edu.}}


\author{    Donghui Yan\\ 
            Department of Statistics\\ University of California, Berkeley, CA 94720 \\~\\
        Aiyou Chen\\
        Google Inc, Mountain View, CA 94043\\\\
        Michael I. Jordan\\
        Department of Statistics and of EECS\\ University of California, Berkeley, CA 94720
}
\date{}
\maketitle

\begin{abstract}
With inspiration from Random Forests (RF) in the context of classification, a
new clustering ensemble method---Cluster Forests (CF) is proposed.
Geometrically, CF randomly probes a high-dimensional data cloud to
obtain ``good local clusterings" and then aggregates via spectral
clustering to obtain cluster assignments for the whole dataset.  The
search for good local clusterings is guided by a cluster quality
measure kappa. CF progressively improves each local clustering in
a fashion that resembles the tree growth in RF. Empirical studies on
several real-world datasets under two different performance metrics
show that CF compares favorably to its competitors.  Theoretical
analysis reveals that the kappa measure makes it possible to grow the
local clustering in a desirable way---it is ``noise-resistant''. A
closed-form expression is obtained for the mis-clustering rate of
spectral clustering under a perturbation model, which yields new
insights into some aspects of spectral clustering.
\end{abstract}





\section{Motivation}
\label{section:introduction}

The general goal of clustering is to partition a set of data such
that data points within the same cluster are ``similar" while those
from different clusters are ``dissimilar." An emerging trend is that
new applications tend to generate data in very high dimensions for
which traditional methodologies of cluster analysis do not work
well. Remedies include dimension reduction and feature
transformation, but it is a challenge to develop effective
instantiations of these remedies in the high-dimensional clustering
setting.  In particular, for datasets whose dimension is beyond
$20$, it is infeasible to perform full subset selection.  Also,
there may not be a single set of attributes on which the whole set
of data can be reasonably separated. Instead, there may be local
patterns in which different choices of attributes or different
projections reveal the clustering.

Our approach to meeting these challenges is to randomly probe the
data/feature space to detect many locally ``good" clusterings and
then aggregate by spectral clustering. The intuition is that, in
high-dimensional spaces, there may be projections or subsets of the
data that are well separated and these projections or subsets may carry
information about the cluster membership of the data involved. If we
can effectively combine such information from many different views
(here a view has two components, the directions or projections we
are looking at and the part of data that are involved), then we can
hope to recover the cluster assignments for the whole dataset.
However, the number of projections or subsets that are potentially
useful tend to be huge, and it is not feasible to conduct a grand
tour of the whole data space by exhaustive search. This motivates us
to randomly probe the data space and then aggregate.

The idea of random projection has been explored in various problem
domains such as clustering \cite{DasguptaRP2000, FernBrodley2003},
manifold learning \cite{HegdeBaraniuk2007} and compressive sensing
\cite{DonohoCS2006}.  However, the most direct motivation for our
work is the Random Forests (RF) methodology for classification
\cite{RF}. In RF, a \emph{bootstrap step} selects a subset of data
while the \emph{tree growth step} progressively improves a tree from
the root downwards---each tree starts from a random collection of
variables at the root and then becomes stronger and stronger as more
nodes are split.  Similarly, we expect that it will be useful in the
context of high-dimensional clustering to go beyond simple random
probings of the data space and to perform a controlled probing in
hope that most of the probings are ``strong." This is achieved by
progressively refining our ``probings" so that eventually each of
them can produce relatively high-quality clusters although they may
start weak.  In addition to the motivation from RF, we note that
similar ideas have been explored in the projection pursuit
literature for regression analysis and density estimation (see
\cite{HuberProjPursuit} and references therein).

RF is a supervised learning methodology and as such there is a clear
goal to achieve, i.e., good classification or regression
performance. In clustering, the goal is less apparent. But
significant progress has been made in recent years in treating
clustering as an optimization problem under an explicitly defined
cost criterion; most notably in the spectral clustering
methodology~\cite{LuxburgSC:2007,ZhangJordan:2008}. Using such
criteria makes it possible to develop an analog of RF in the
clustering domain.

Our contributions can be summarized as follows. We propose a new
cluster ensemble method that incorporates model selection and
regularization. Empirically CF compares favorably to some popular
cluster ensemble methods as well as spectral clustering. The
improvement of CF over the base clustering algorithm ($K$-means
clustering in our current implementation) is substantial. We also
provide some theoretical support for our work: (1) Under a
simplified model, CF is shown to grow the clustering instances in a
``noise-resistant'' manner; (2) we obtain a closed-form formula for
the mis-clustering rate of spectral clustering under a perturbation
model that yields new insights into aspects of spectral clustering
that are relevant to CF.

The remainder of the paper is organized as follows. In
Section~\ref{section:method}, we present a detailed description of
CF. Related work is discussed in Section~\ref{section:related}. This
is followed by an analysis of the $\kappa$ criterion and the
mis-clustering rate of spectral clustering under a perturbation
model in Section~\ref{section:littleTheory}.  We evaluate our method
in Section~\ref{section:experiment} by simulations on Gaussian
mixtures and comparison to several popular cluster ensemble methods
as well as spectral clustering on some UC Irvine machine learning
benchmark datasets. Finally we conclude in
Section~\ref{section:conclusion}.

\section{The Method}
\label{section:method}

CF is an instance of the general class of \emph{cluster ensemble
methods}~\cite{StrehlGhosh2002}, and as such it is comprised of two
phases: one which creates many cluster instances and one which
aggregates these instances into an overall clustering.  We begin by
discussing the cluster creation phase.

\subsection{Growth of clustering vectors}
\label{section:clusteringTree}

CF works by aggregating many instances of clustering problems, with
each instance based on a different subset of features (with varying
weights).  We define the \emph{feature space}
$\mathcal{F}=\{1,2, \ldots, p\}$ as the set of indices of coordinates in
$\mathbb{R}^p$.  We assume that we are given $n$ i.i.d.\
observations $X_1, \ldots, X_n \in \mathbb{R}^p$. A \emph{clustering
vector} is defined to be a subset of the feature space.

\textbf{Definition 1.} The growth of a clustering vector is governed
by the following cluster quality measure:
\begin{equation}
\label{eq:kappaDef}
\kappa(\bm{\tilde{f}})=\frac{SS_W(\bm{\tilde{f}})}{SS_B(\bm{\tilde{f}})},
\end{equation}
where $SS_W$ and $SS_B$ are the within-cluster and between-cluster
sum of squared distances (see Section~\ref{section:proofThm1}),
computed on the set of features currently in use (denoted by
$\bm{\tilde{f}}$), respectively.

Using the quality measure $\kappa$, we iteratively expand the clustering
vector. Specifically, letting $\tau$ denote the number of
consecutive unsuccessful attempts in expanding the clustering vector
$\bm{\tilde{f}}$, and letting $\tau_m$ be the maximal allowed value
of $\tau$, the growth of a clustering vector is described in
Algorithm~\ref{algorithm:clusteringVector}.

\begin{algorithm}
\caption{~The growth of a clustering vector $\bm{\tilde{f}}$}
\label{algorithm:clusteringVector}
\begin{algorithmic}[1]
\STATE Initialize $\bm{\tilde{f}}$ to be NULL and set $\tau=0$;
\STATE Apply feature competition and update $\bm{\tilde{f}}
\leftarrow (f_1^{(0)}, \ldots, f_b^{(0)})$; 
\REPEAT 
\STATE Sample $b$ features, denoted as $f_1, \ldots, f_b$, from the
feature space $\mathcal{F}$; 
\STATE Apply $K$-means (the {\it base clustering algorithm}) to the
data induced by the feature vector $(\bm{\tilde{f}}, f_1, \ldots, f_b)$;
\IF{$\kappa(\bm{\tilde{f}},f_1 \ldots, f_b) < \kappa(\bm{\tilde{f}})$} 
\STATE expand $\bm{\tilde{f}}$ by $\bm{\tilde{f}} \leftarrow
(\bm{\tilde{f}}, f_1, \ldots, f_b)$ and set $\tau \gets 0$; 
\ELSE \STATE discard $\{f_1, \ldots, f_b\}$ and set $\tau \gets \tau+1$;
\ENDIF 
\UNTIL{$\tau \geq \tau_m$}.
\end{algorithmic}
\end{algorithm}

\begin{algorithm}
\caption{~Feature competition} \label{algorithm:featureCompetition}
\begin{algorithmic}[1]
\FOR {$i=1$ $\bf{to}$ $q$} 
\STATE Sample $b$ features, $f_1^{(i)}, \ldots, f_b^{(i)}$, from the
feature space $\mathcal{F}$; 
\STATE Apply $K$-means to the data projected on
$(f_1^{(i)}, \ldots, f_b^{(i)})$ to get
$\kappa(f_1^{(i)}, \ldots, f_b^{(i)})$; 
\ENDFOR 
\STATE Set $(f_1^{(0)}, \ldots, f_b^{(0)}) \leftarrow \arg\min_{i=1}^q
\kappa(f_1^{(i)}, \ldots, f_b^{(i)})$.
\end{algorithmic}
\end{algorithm}

Step 2 in Algorithm~\ref{algorithm:clusteringVector} is called
\emph{feature competition} (setting $q=1$ reduces to the usual
mode). It aims to provide a good initialization for the growth of a
clustering vector. The feature competition procedure is detailed in
Algorithm~\ref{algorithm:featureCompetition}.

Feature competition is motivated by
Theorem~\ref{theorem:noiseResistent} in
Section~\ref{section:kappaCriterion}---it helps prevent noisy or
``weak" features from entering the clustering vector at the
initialization, and, by Theorem~\ref{theorem:noiseResistent}, the
resulting clustering vector will be formed by ``strong" features
which can lead to a ``good" clustering instance. This will be
especially helpful when the number of noisy or very weak features is
large. Note that feature competition can also be applied in other
steps in growing the clustering vector. A heuristic for choosing $q$
is based on the ``feature profile plot,'' a detailed discussion of
which is provided in Section~\ref{section:expUCI}.

\subsection{The CF algorithm}

The CF algorithm is detailed in Algorithm~\ref{algorithm:CF}. The
key steps are: (a) grow $T$ clustering vectors and obtain the
corresponding clusterings; (b) average the clustering matrices to
yield an aggregate matrix $P$; (c) regularize $P$; and (d) perform
spectral clustering on the regularized matrix. The regularization
step is done by {\it thresholding} $P$ at level $\beta_2$; that is,
setting $P_{ij}$ to be $0$ if it is less than a constant $\beta_2
\in (0,1)$, followed by a further nonlinear transformation $P
\leftarrow \exp(\beta_1 P)$ which we call {\it scaling}.

\begin{algorithm}
\caption{~The CF algorithm} \label{algorithm:CF}
\begin{algorithmic}[1]
\FOR {$l=1$ $\bf{to}$ $T$} 
\STATE Grow a clustering vector, $\bm{\tilde{f}}^{(l)}$, according
to Algorithm~\ref{algorithm:clusteringVector}; 
\STATE Apply the base clustering algorithm to the data induced by
clustering vector $\bm{\tilde{f}}^{(l)}$ to get a partition of the
data;
\STATE Construct $n \times n$ co-cluster indicator matrix (or
affinity matrix) $P^{(l)}$
\begin{equation*}
P_{ij}^{(l)}=\left\{
\begin{array}{ll}
1, & ~\mbox{if}~X_i~\mbox{and}~X_j~\mbox{are in the same cluster} \\
0, & ~\mbox{otherwise};
\end{array}\right.
\end{equation*}
\ENDFOR
\STATE Average the indicator matrices to get $P \leftarrow
\frac{1}{T} \sum_{l=1}^T P^{(l)}$; 
\STATE Regularize the matrix $P$; 
\STATE Apply spectral clustering to $P$ to get the final clustering.
\end{algorithmic}
\end{algorithm}

We provide some justification for our choice of spectral clustering
in Section~\ref{XXX}. As the entries of matrix $P$ can be viewed as
encoding the pairwise similarities between data points (each
$P^{(l)}$ is a positive semidefinite matrix and the average matrix
$P$ is thus positive semidefinite and a valid kernel matrix), any clustering algorithm
based on pairwise similarity can be used as the aggregation engine.
Throughout this paper, we use normalized cuts (Ncut,
\cite{ShiMalik2000}) for spectral clustering.

\section{Related Work} \label{section:related} In this section, we
compare and contrast CF to other work on cluster ensembles. It is
beyond our scope to attempt a comprehensive review of the enormous
body of work on clustering, please refer to \cite{JainMurty1999,
HTF2009} for overview and references.  We will also omit a
discussion on classifier ensembles, see \cite{HTF2009} for
references. Our focus will be on cluster ensembles.  We discuss the
two phases of cluster ensembles, namely, the generation of multiple
clustering instances and their aggregation, separately.

For the generation of clustering instances, there are two main
approaches---data re-sampling and random projection.
\cite{DudoitFridlyand2003} and \cite{MinaeiTopchy2004} produce
clustering instances on bootstrap samples from the original data.
Random projection is used by \cite{FernBrodley2003} where each
clustering instance is generated by randomly projecting the data to
a lower-dimensional subspace. These methods are myopic in that they
do not attempt to use the quality of the resulting clusterings to
choose samples or projections. Moreover, in the case of random
projections, the choice of the dimension of the subspace is myopic.
In contrast, CF proceeds by selecting features that progressively
improve the quality (measured by $\kappa$) of individual clustering
instances in a fashion resembling that of RF. As individual
clustering instances are refined, better final clustering
performance can be expected.  We view this non-myopic approach to
generating clustering instances as essential when the data lie in a
high-dimensional ambient space.  Another possible approach is to
generate clustering instances via random restarts of a base
clustering algorithm such as $K$-means~\cite{FredJain2002}.

The main approaches to aggregation of clustering instances are the
\emph{co-association method} \cite{StrehlGhosh2002,FredJain2005}
and the \emph{hyper-graph method} \cite{StrehlGhosh2002}. The
co-association method counts the number of times two points fall in
the same cluster in the ensemble. The hyper-graph method solves a
$k$-way minimal cut hyper-graph partitioning problem where a vertex
corresponds to a data point and a link is added between two vertices
each time the two points meet in the same cluster. Another approach
is due to \cite{TopchyJain2003}, who propose to combine clustering
instances with mixture modeling where the final clustering is
identified as a maximum likelihood solution.  CF is based on
co-association, specifically using spectral clustering for
aggregation. Additionally, CF incorporates regularization such that
the pairwise similarity entries that are close to zero are
thresholded to zero.
This yields improved clusterings as demonstrated by our empirical
studies.

A different but closely related problem is clustering aggregation
\cite{GionisMannila2005} which requires finding a clustering that
``agrees" as much as possible with a given set of input clustering
instances. Here these clustering instances are assumed to be known
and the problem can be viewed as the second stage of clustering
ensemble. Also related is ensemble selection \cite{CaruanaNCK2004,
FernLin2008, AzimiFern2009} which is applicable to CF but this is
not the focus of the present work. Finally there is unsupervised
learning with random forests \cite{RF,ShiHorvath2006} where RF is
used for deriving a suitable distance metric (by synthesizing a copy
of the data via randomization and using it as the ``contrast"
pattern); this methodology is fundamentally different from ours.

\section{Theoretical Analysis}
\label{section:littleTheory} In this section, we provide a
theoretical analysis of some aspects of CF.  In particular we
develop theory for the $\kappa$ criterion, presenting conditions
under which CF is ``noise-resistant." By ``noise-resistant" we mean
that the algorithm can prevent a pure noise feature from entering
the clustering vector.  We also present a perturbation analysis of
spectral clustering, deriving a closed-form expression for the
mis-clustering rate.

\subsection{CF with $\kappa$ is noise-resistant}
\label{section:kappaCriterion} We analyze the case in which the
clusters are generated by a Gaussian mixture:
\begin{eqnarray}
\label{eq:gm} \Delta\mathcal{N}(\bm{\mu},\Sigma) +(1-\Delta)
\mathcal{N}(-\bm{\mu},\Sigma),
\end{eqnarray}
where $\Delta \in \{0,1\}$ with $\mathbb{P}(\Delta=1)=\pi$ specifies
the cluster membership of an observation, and
$\mathcal{N}(\bm{\mu},\Sigma)$ stands for a Gaussian random variable
with mean $\bm{\mu}=(\mu[1], \ldots, \mu[p]) \in \mathbb{R}^p$ and
covariance matrix $\Sigma$.  We specifically consider
$\pi=\frac{1}{2}$ and $\Sigma=\bm{I}_{p \times p}$; this is a simple
case which yields some insight into the feature selection ability of
CF.  We start with a few definitions.

\textbf{Definition 2.} Let $h: \mathbb{R}^p \mapsto \{0,1\}$ be a
{\it decision rule}. Let $\Delta$ be the cluster membership for
observation $X$. A loss function associated with $h$ is defined as
\begin{equation}
\label{def:lossFunction} l(h(X),\Delta)=\left\{
\begin{array}{cl}
0, & ~\mbox{if}~h(X)=\Delta \\
1, & ~\mbox{otherwise}.
\end{array}\right.
\end{equation}
The optimal clustering rule under \eqref{def:lossFunction} is
defined as
\begin{equation}
\label{def:optimalRule} h^*=\arg\min_{h \in \mathcal{G}} \mathbb{E}
l(h(X),\Delta),
\end{equation}
where $\mathcal{G} \triangleq \left\{h:~\mathbb{R}^p \mapsto
\{0,1\}\right\}$ and the expectation is taken with respect to the
random vector $(X,\Delta)$.

\textbf{Definition 3 \cite{Pollard1981}.} For a probability measure
$Q$ on $\mathbb{R}^d$ and a finite set $A \subseteq \mathbb{R}^d$,
define {\it the within cluster sum of distances} by
\begin{equation}
\label{eq:sswDef} \Phi(A,Q)=\int \min_{a \in A} \phi(||x-a||)Q(dx),
\end{equation}
where $\phi(||x-a||)$ defines the distance between points $x$ and $a
\in A$. $K$-means clustering seeks to minimize $\Phi(A,Q)$ over a
set $A$ with at most $K$ elements.  We focus on the case
$\phi(x)=x^2$, $K=2$ and refer to \[\{\mu_0^*, \mu_1^*\}
=\arg\min_{A} ~\Phi(A,Q)\] as the {\it population cluster centers}.

\textbf{Definition 4.} The $i^{th}$ feature is called a {\it noise}
feature if $\mu[i]=0$ where $\mu[i]$ denotes the $i^{th}$ coordinate
of $\bm{\mu}$. A feature is {\it ``strong" (``weak")} if $|\mu[i]|$
is ``large" (``small").

\begin{theorem}
\label{theorem:noiseResistent} Assume the cluster is generated by
Gaussian mixture \eqref{eq:gm} with $\Sigma=\bm{I}$ and
$\pi=\frac{1}{2}$. Assume one feature is considered at each step and
duplicate features are excluded. Let $\mathcal{I} \neq \emptyset$ be
the set of features currently in the clustering vector and let $f_n$
be a noise feature such that $f_n \notin \mathcal{I}$. If $\sum_{i
\in \mathcal{I}} \left(\mu_0^*[i] - \mu_1^*[i]\right)^2>0$, then
$\kappa(\mathcal{I}) < \kappa(\{\mathcal{I},f_n\})$.
\end{theorem}

\textbf{Remark.} The interpretation of
Theorem~\ref{theorem:noiseResistent} is that noise features are
generally not included in cluster vectors under the CF procedure;
thus, CF with the $\kappa$ criterion is noise-resistant.

The proof of Theorem~\ref{theorem:noiseResistent} is in the
appendix. It proceeds by explicitly calculating $SS_B$ and $SS_W$
(see Section~\ref{section:proofThm1}) and thus an expression for
$\kappa=SS_W/ SS_B$. The calculation is facilitated by the
equivalence, under $\pi=\frac{1}{2}$ and $\Sigma=\bm{I}$, of
$K$-means clustering and the optimal clustering rule $h^*$ under
loss function \eqref{def:lossFunction}.

\subsection{Quantifying the mis-clustering rate}
\label{XXX}

Recall that spectral clustering works on a weighted similarity graph
$\mathcal{G}(\bm{V},P)$ where $\bm{V}$ is formed by a set of data
points, $X_i, i=1, \ldots, n$, and $P$ encodes their pairwise similarities.
Spectral clustering algorithms compute the eigendecomposition of the
Laplacian matrix (often symmetrically normalized as $\mathcal{L}(P)=D^{-1/2}(I-P)D^{-1/2}$
where $D$ is a diagonal matrix with diagonals being degrees of $\mathcal{G}$).
Different notions of similarity and ways of using the spectral
decomposition lead to different spectral cluster algorithms
\cite{ShiMalik2000,MeilaShi2001, NgJordan2002,KannanVempala2004,
LuxburgSC:2007,ZhangJordan:2008}.  In particular, Ncut
\cite{ShiMalik2000} forms a bipartition of the data according to
the sign of the components of the second eigenvector (i.e.,
corresponding to the second smallest eigenvalue) of
$\mathcal{L}(P)$. On each of the two partitions, Ncut is then
applied recursively until a stopping criterion is met.

There has been relatively little theoretical work on spectral
clustering; exceptions include \cite{AzarFiat2001, NgJordan2002,
KannanVempala2004, VempalaWang2004, NadlerDiffusion2005,
AchlioptasMcSherry2005, LuxburgBelkin2008}. Here we analyze the
mis-clustering rate for symmetrically normalized spectral
clustering. For simplicity we consider the case of two clusters
under a perturbation model.

\begin{figure}[H]
\centering
\begin{center}
\hspace{0cm}
\includegraphics*[scale=0.6,clip]{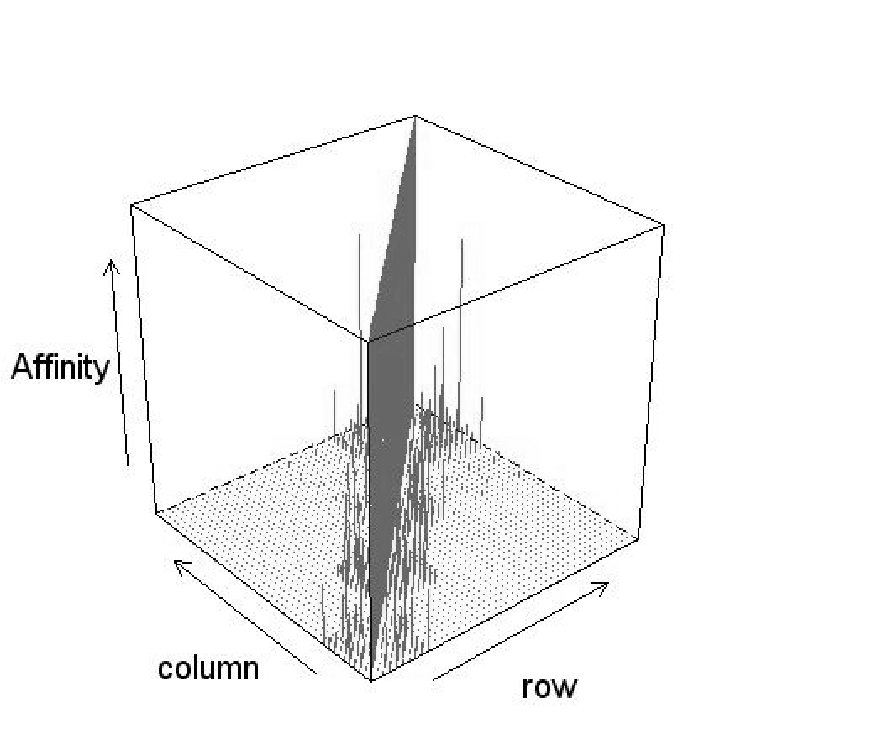}
\end{center}
\caption{The affinity matrix produced by CF for the Soybean dataset
with $4$ clusters. The number of clustering vectors in the ensemble
is $100$.} \label{figure:affMatSoybean}
\end{figure}

Assume that the similarity (affinity) matrix can be written as
\begin{equation}
\label{eq:perturbModel} P=\overline{P}+\varepsilon,
\end{equation}
where
\begin{equation*}
\overline{P}_{ij}=\left\{ \begin{array}{ll}
1-\nu, & ~\mbox{if}~i,j\leq n_{1}~\text{or}~i,j>n_{1}\vspace{1pt}\\
\nu, & ~\mbox{otherwise},\end{array}\right.
\end{equation*}
and $\varepsilon=(\varepsilon_{ij})_1^n$ is a symmetric random
matrix with $\mathbb{E}\varepsilon_{ij}=0$. Here $n_1$ and
$n_2$ are the size of the two clusters. Let $n_{2}=\gamma n_{1}$ and
$n=n_{1}+n_{2}$. Without loss of generality, assume $\gamma \leq 1$.
Similar models have been studied in earlier work; see, for instance,
\cite{HollandLL1983, NowickiSnijders2001, AiroldiBFX2008,
BickelChen2009}. Our focus is different; we aim at the
mis-clustering rate due to perturbation. Such a model is appropriate
for modeling the similarity (affinity) matrix produced by CF. For
example, Figure~\ref{figure:affMatSoybean} shows the affinity matrix
produced by CF on the Soybean data \cite{UCI}; this matrix is nearly
block-diagonal with each block corresponding to data points from the
same cluster (there are four of them in total) and the off-diagonal
elements are mostly close to zero. Thus a perturbation model such as
\eqref{eq:perturbModel} is a good approximation to the similarity
matrix produced by CF and can potentially allow us to gain insights
into the nature of CF.

Let $\mathcal{M}$ be the mis-clustering rate, i.e., the proportion
of data points assigned to a wrong cluster (i.e., $h(X) \neq
\Delta$). Theorem~\ref{thm:Ncut} characterizes the expected value of
$\mathcal{M}$ under perturbation model \eqref{eq:perturbModel}.

\begin{theorem}\label{thm:Ncut} Assume that $\varepsilon_{ij}$,
$i\geq j$ are mutually independent $\mathcal{N}(0,\sigma^{2})$. Let
$0<\nu \ll \gamma \leq 1$. Then
\begin{equation}
\label{eq:M} \lim_{n \rightarrow \infty}
\frac{1}{n}\log(\mathbb{E}\mathcal{M})=
-\frac{\gamma^{2}}{2\sigma^2(1+\gamma)(1+\gamma^{3})}.
\end{equation}
\end{theorem}

The proof is in the appendix. The main step is to obtain an analytic
expression for the second eigenvector of $\mathcal{L}(P)$. Our
approach is based on matrix perturbation theory \cite{Kato1966},
and the key idea is as follows.

Let $\mathfrak{P}(A)$ denote the eigenprojection of a linear
operator $A: \mathbb{R}^n \rightarrow \mathbb{R}^n$. Then,
$\mathfrak{P}(A)$ can be expressed explicitly as the following
contour integral
\begin{equation}
\label{eq:contourIntegral} \mathfrak{P}(A) = \frac{1}{2\pi
i}\oint_{\Gamma} (A-\zeta I)^{-1}d\zeta,
\end{equation}
where $\Gamma$ is a simple Jordan curve enclosing the eigenvalues of
interest (i.e., the first two eigenvalues of matrix
$\mathcal{L}(P)$) and excluding all others. The eigenvectors of
interest can then be obtained by
\begin{equation}
\label{eq:eigenprojectionVector} \varphi_i=\mathfrak{P}\omega_i,
\quad i=1,2,
\end{equation}
where $\omega_i, i=1,2$ are fixed linearly independent vectors in
$\mathbb{R}^n$. An explicit expression for the second eigenvector
can then be obtained under perturbation model
\eqref{eq:perturbModel}, which we use to calculate the final
mis-clustering rate.

\textbf{Remarks.} While formula~\eqref{eq:M} is obtained under some
simplifying assumptions, it provides insights into the nature of
spectral clustering.
\begin{enumerate}
\item [1).] The mis-clustering rate increases as $\sigma$ increases.
\item [2).]
By checking the derivative, the right-hand side of \eqref{eq:M} can be
seen to be a unimodal function of $\gamma$, minimized at $\gamma =1$
with a fixed $\sigma$. Thus the mis-clustering rate decreases as the
cluster sizes become more balanced.
\item [3).] When $\gamma$ is very small, i.e., the clusters are extremely unbalanced,
spectral clustering is likely to fail.
\end{enumerate}
These results are consistent with existing empirical findings. In
particular, they underscore the important role played by the ratio
of two cluster sizes, $\gamma$, on the mis-clustering rate.
Additionally, our analysis (in the proof of Theorem~\ref{thm:Ncut})
also implies that the best cutoff value (when assigning cluster
membership based on the second eigenvector) is not exactly zero but
shifts slightly towards the center of those components of the second
eigenvector that correspond to the smaller cluster.  Related work
has been presented by \cite{HuangYanNips2008f} who study an end-to-end
perturbation yielding a final mis-clustering rate that is \emph{approximate}
in nature.  Theorem~\ref{thm:Ncut} is based on a perturbation model
for the affinity matrix and provides, for the first time, a
closed-form expression for the mis-clustering rate of spectral
clustering under such a model.

\section{Experiments}
\label{section:experiment} We present results from two sets of
experiments, one on synthetic data, specifically designed to demonstrate the
feature selection and ``noise-resistance" capability of CF, and the
other on several real-world datasets \cite{UCI} where we compare
the overall clustering performance of CF with several competitors,
as well as spectral clustering, under two different metrics. These
experiments are presented in separate subsections.

\subsection{Feature selection capability of CF}
In this subsection, we describe three simulations that aim to study
the feature selection capability and ``noise-resistance" feature of
CF. Assume the underlying data are generated i.i.d.\ by Gaussian
mixture \eqref{eq:gm}.

\begin{figure}[H]
\centering
\begin{center}
\hspace{-0.1in}
\includegraphics*[scale=0.25,clip]{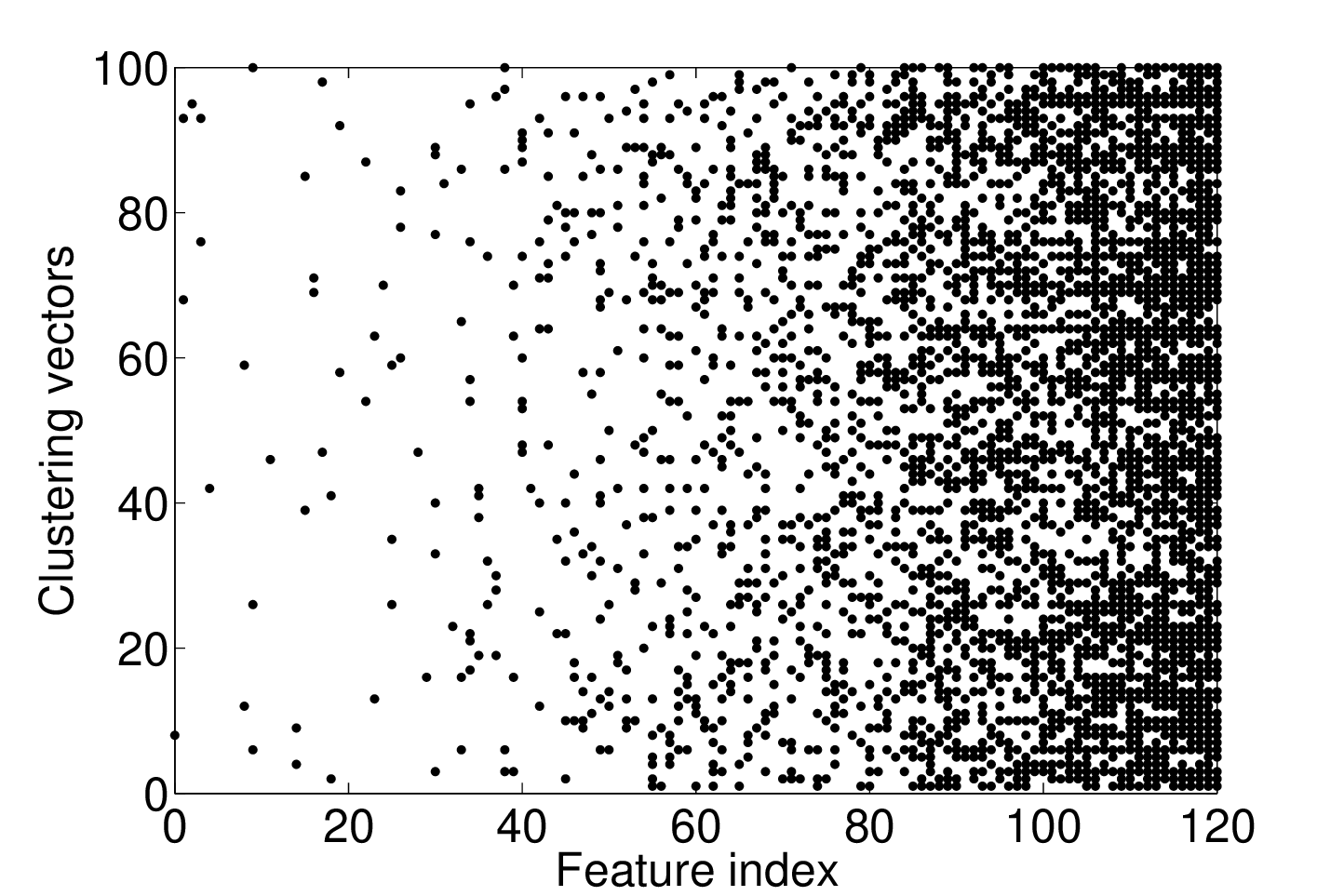}\hspace{-0.05in}
\includegraphics*[scale=0.25,clip]{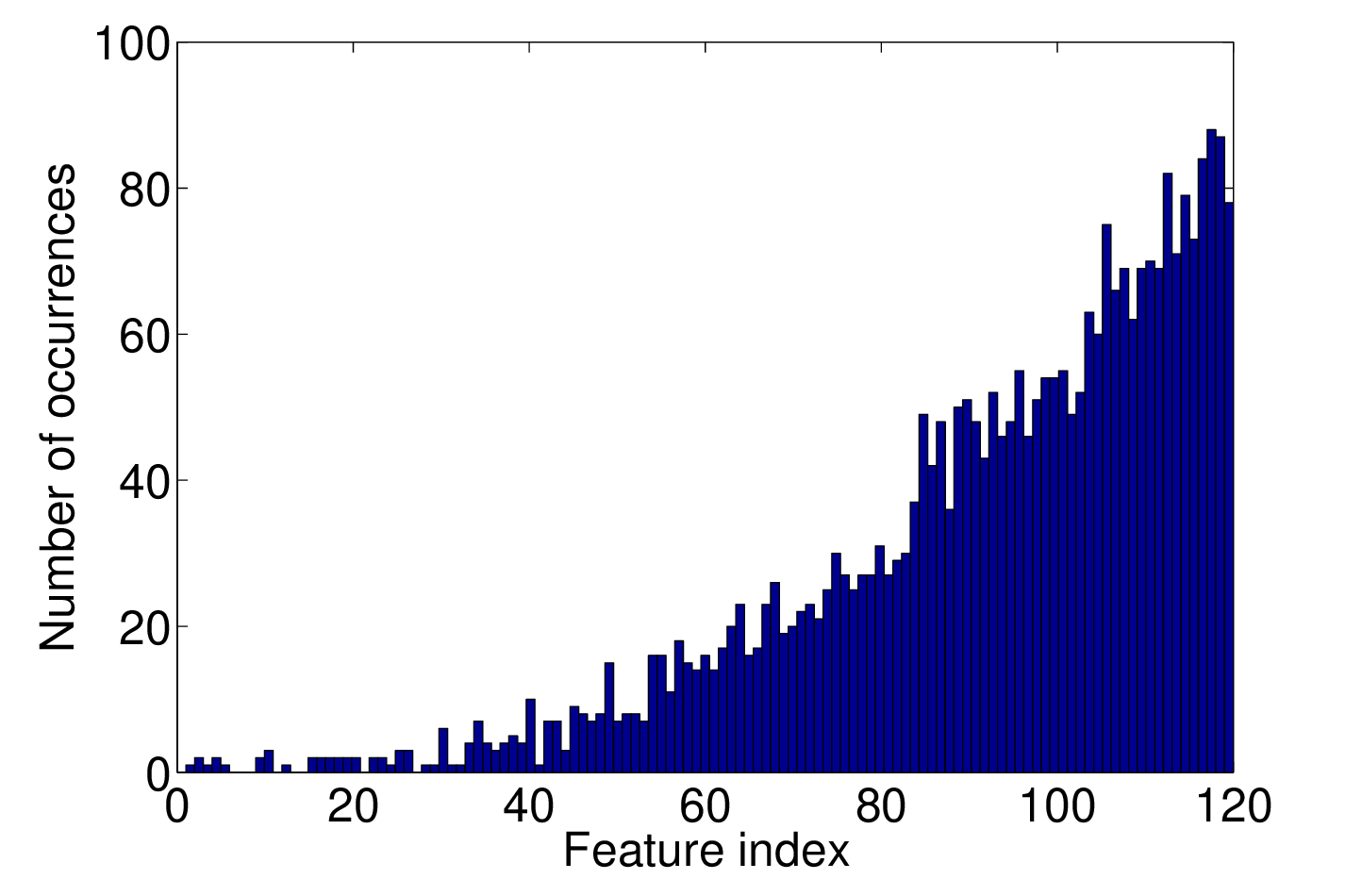}
\end{center}
\caption{The occurrence of individual features in the $100$
clustering vectors for $\bm{G}_1$. The left plot shows the features
included (indicated by a solid circle) in each clustering vector.
Each horizontal line corresponds to a clustering vector. The right
plot shows the total number of occurrences of each feature.}
\label{figure:clusterVectors}
\end{figure}

In the first simulation, a sample of $4000$ observations is
generated from \eqref{eq:gm} with $\bm{\mu}=(0, \ldots, 0,1,2, \ldots, 100)^T$
and the diagonals of $\Sigma$ are all $1$ while the non-diagonals
are i.i.d.\ uniform from $[0,0.5]$ subject to symmetry and positive
definitiveness of $\Sigma$. Denote this dataset as $\bm{G}_1$. At
each step of cluster growing one feature is sampled from
$\mathcal{F}$ and tested to see if it is to be included in the
clustering vector by the $\kappa$ criterion. We run the clustering
vector growth procedure until all features have been attempted with
duplicate features excluded. We generate $100$ clustering vectors
using this procedure.  In Figure~\ref{figure:clusterVectors},
all but one of the $100$ clustering vectors include at least one
feature from the top three features (ranked according to the $|\mu[i]|$
value) and all clustering vectors contain at least one of the top
five features.
\begin{figure}[H]
\centering
\begin{center}
\includegraphics*[scale=0.26,clip]{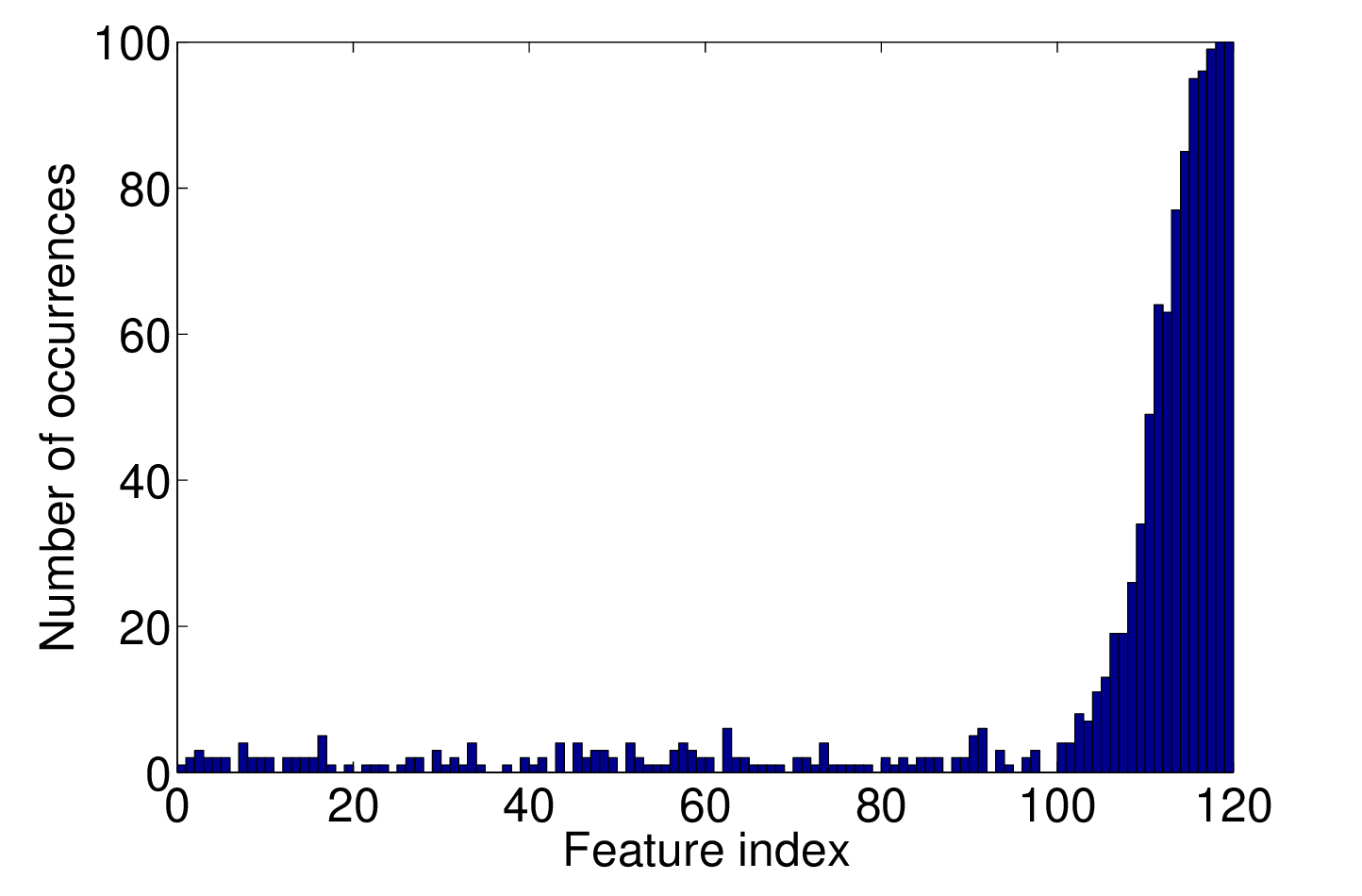}
\includegraphics*[scale=0.26,clip]{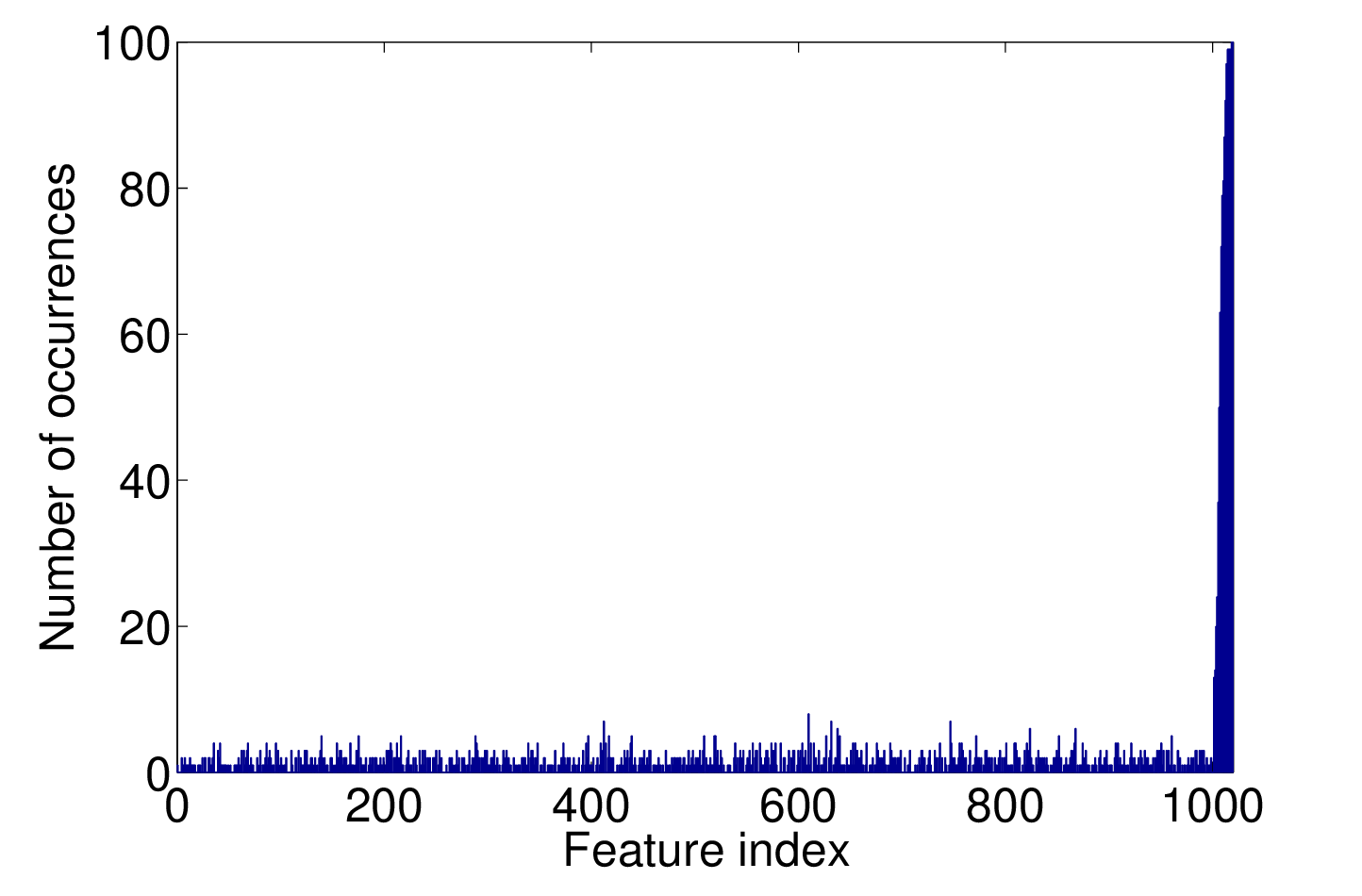}
\end{center}
\caption{The occurrence of individual features in the $100$
clustering vectors for $\bm{G}_2$ and $\bm{G}_3$. The left plot is
for $\bm{G}_2$ where the first $100$ features are noise features.
The right plot is for $\bm{G}_3$ where the first $1000$ features are
noisy features.} \label{figure:clusterVectors2}
\end{figure}
\normalsize

We also performed a simulation with ``noisy" data. In this
simulation, data are generated from \eqref{eq:gm} with
$\Sigma=\bm{I}$, the identity matrix, such that the first $100$
coordinates of $\bm{\mu}$ are $0$ and the next $20$ are generated
i.i.d.\ uniformly from $[0,1]$. We denote this dataset as $\bm{G}_2$.
Finally, we also considered an extreme case where data are generated
from \eqref{eq:gm} with $\Sigma=\bm{I}$ such that the first $1000$
features are noise features and the remaining $20$ are useful
features (with coordinates of $\mu$ from $\pm 1$ to $\pm 20$); this
is denoted as $\bm{G}_3$. The occurrences of individual features for
$\bm{G}_2$ and $\bm{G}_3$ are shown in
Figure~\ref{figure:clusterVectors2}. Note that the two plots in
Figure~\ref{figure:clusterVectors2} are produced by invoking feature
competition with $q=20$ and $q=50$, respectively. It is worthwhile
to note that, for both $\bm{G}_2$ and $\bm{G}_3$, despite the fact
that a majority of features are pure noise ($100$ out of a total of
$120$ for $\bm{G}_2$ or $1000$ out of $1020$ for $\bm{G}_3$,
respectively), CF achieves clustering accuracies (computed against
the true labels) that are very close to the Bayes rates (about $1$).
\subsection{Experiments on UC Irvine datasets}
\label{section:expUCI} We conducted experiments on eight UC Irvine
datasets~\cite{UCI}, the Soybean, SPECT Heart, image segmentation
(ImgSeg), Heart, Wine, Wisconsin breast cancer (WDBC), robot execution
failures (lp5, Robot) and the
Madelon datasets. A summary is provided in
Table~\ref{table:datasets}. It is interesting to note that the
Madelon dataset has only $20$ useful features out of a total of
$500$ (but such information is not used in our experiments). Note
that true labels are available for all eight datasets. We use the
labels to evaluate the performance of the clustering methods, while
recognizing that this evaluation is only partially satisfactory. Two
different performance metrics, $\rho_r$ and $\rho_c$, are used in
our experiments.

\begin{table}[H]
\begin{center}
\begin{tabular}{r|ccc}
\hline
Dataset          & Features     &  Classes   &  \#Instances\\
\hline
Soybean          & 35              & 4             & 47\\
SPECT            & 22              & 2             & 267\\
ImgSeg           & 19              & 7             & 2100\\
Heart            & 13              & 2             & 270\\
Wine             & 13              & 3             & 178\\
WDBC             & 30              & 2             & 569\\
Robot            & 90              & 5             & 164\\
Madelon          & 500             & 2             & 2000 \\
\hline
\end{tabular}
\caption{\small A summary of datasets.} \label{table:datasets}
\end{center}
\end{table}

\textbf{Definition 5.} One measure of the quality of a cluster
ensemble is given by
\begin{equation*}
\rho_r=\frac{\mbox{Number of correctly clustered pairs}}{\mbox{Total
number of pairs}},
\end{equation*}
where by ``correctly clustered pair'' we mean two instances have the
same co-cluster membership (that is, they are in the same cluster)
under both CF and the labels provided in the original dataset. 

\textbf{Definition 6.} Another performance metric is the clustering
accuracy. Let $z=\{1,2, \ldots, J\}$ denote the set of class labels, and
$\theta(.)$ and $f(.)$ the true label and the label obtained by a
clustering algorithm, respectively. The clustering accuracy is
defined as
\begin{equation}
\label{clusterAccuracy} \rho_c(f)=\max_{\tau \in \Pi_{\bm{z}}}
\left\{\frac{1}{n}\sum_{i=1}^n \mathbb{I}\{\tau
\left(f(X_i)\right)=\theta(X_i)\}\right\},
\end{equation}
where $\mathbb{I}$ is the indicator function and $\Pi_{\bm{z}}$ is
the set of all permutations on the label set $\bm{z}$. This measure
is a natural extension of the classification accuracy (under 0-1
loss) and has been used by a number of authors, e.g.,
\cite{MeilaShortreed2005, YanHuangJordan2009}.

The idea of having two different performance metrics is to assess a
clustering algorithm from different perspectives since one metric
may particularly favor certain aspects while overlooking others. For
example, in our experiment we observe that, for some datasets, some
clustering algorithms (e.g., RP or EA) achieve a high value of
$\rho_r$ but a small $\rho_c$ on the same clustering instance (note
that, for RP and EA on the same dataset, $\rho_c$ and $\rho_r$ as
reported here may be calculated under different parameter settings,
e.g., $\rho_c$ may be calculated when the threshold value $t=0.3$
while $\rho_r$ calculated when $t=0.4$ on a certain dataset).

We compare CF to three other cluster ensemble algorithms---bagged
clustering (bC2, \cite{DudoitFridlyand2003}), random projection (RP,
\cite{FernBrodley2003}), and evidence accumulation (EA,
\cite{FredJain2002}).  We made slight modifications to the original
implementations to standardize our comparison. These include
adopting $K$-means clustering ($K$-medoids is used for bC2 in
\cite{DudoitFridlyand2003} but differs very little from $K$-means on
the datasets we have tried) to be the base clustering algorithm,
changing the agglomerative algorithm used in RP to be based on
single linkage in order to match the implementation in EA.
Throughout we run $K$-means clustering with the R project package
{\it kmeans()} using the ``Hartigan-Wong" algorithm
(\cite{HartiganWong}). Unless otherwise specified, the two
parameters $(n_{it}, n_{rst})$, which stands for the maximum number
of iterations and the number of restarts during each run of {\it
kmeans()}, respectively, are set to be $(200, 20)$.

We now list the parameters used in our implementation. Define the
number of initial clusters, $n_b$, to be that of clusters in running
the base clustering algorithm; denote the number of final clusters
(i.e., the number of clusters provided in the data or ground truth)
by $n_f$. In CF, the scaling parameter $\beta_1$ is set to be $10$
(i.e., $0.1$ times the ensemble size); the thresholding level
$\beta_2$ is 0.4 (we find very little difference in performance by
setting $\beta_2 \in [0.3, 0.5]$); the number of features, $b$,
sampled each time in growing a clustering vector is $2$; we set
$\tau_m=3$ and $n_b=n_f$. (It is possible to vary $n_b$ for gain in
performance, see discussion at the end of this subsection).
Empirically, we find results not particularly sensitive to the
choice of $\tau_m$ as long as $\tau_m \ge 3$. In RP, the search
for the dimension of the target subspace for random projection
is conducted starting from a value of five and proceeding
upwards.  We set $n_b=n_f$. EA \cite{FredJain2002} suggests using
$\sqrt{n}$ ($n$ being the sample size) for $n_b$. This sometimes
leads to unsatisfactory results (which is the case for all except
two of the datasets) and if that happens we replace it with $n_f$.
In EA, the threshold value, $t$, for the single linkage algorithm is
searched through $\{0.3, 0.4, 0.5, 0.6, 0.7, 0.75\}$ as suggested by
\cite{FredJain2002}. In bC2, we set $n_b=n_f$ according to
\cite{DudoitFridlyand2003}.

\begin{table}[H]
\begin{center}
\begin{tabular}{r|cccc}
\hline
Dataset     &CF              &RP       &bC2          &EA        \\
\hline
Soybean     &\bf{92.36}           &87.04    &83.16        &86.48\\
SPECT       &\bf{56.78}           &49.89    &50.61        &51.04\\
ImgSeg      &79.71           &\bf{85.88}    &82.19        &85.75\\
Heart       &\bf{56.90}           &52.41    &51.50        &53.20\\
Wine        &\bf{79.70}           &71.94    &71.97        &71.86\\
WDBC        &\bf{79.66}           &74.89    &74.87        &75.04\\
Robot       &\bf{63.42}           &41.52    &39.76          &58.31\\
Madelon     &50.76             &\bf{50.82}    &49.98        &49.98\\
\hline
\end{tabular}
\caption{$\rho_r$ for different datasets and methods (CF calculated
when $q=1$).} \label{table:simulationResultsR}
\end{center}
\end{table}

\begin{figure}[H]
\centering
\begin{center}
\includegraphics*[scale=0.33,angle=0,clip]{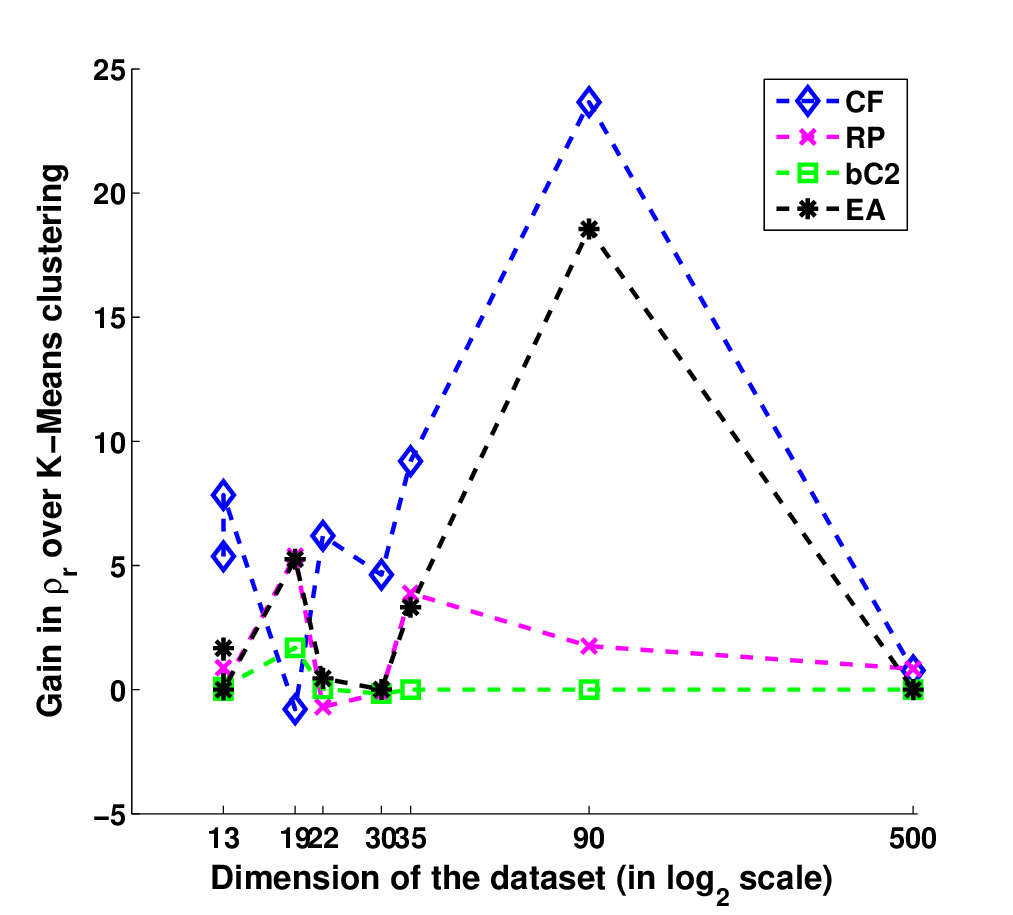}\quad\quad
\includegraphics*[scale=0.33,angle=0,clip]{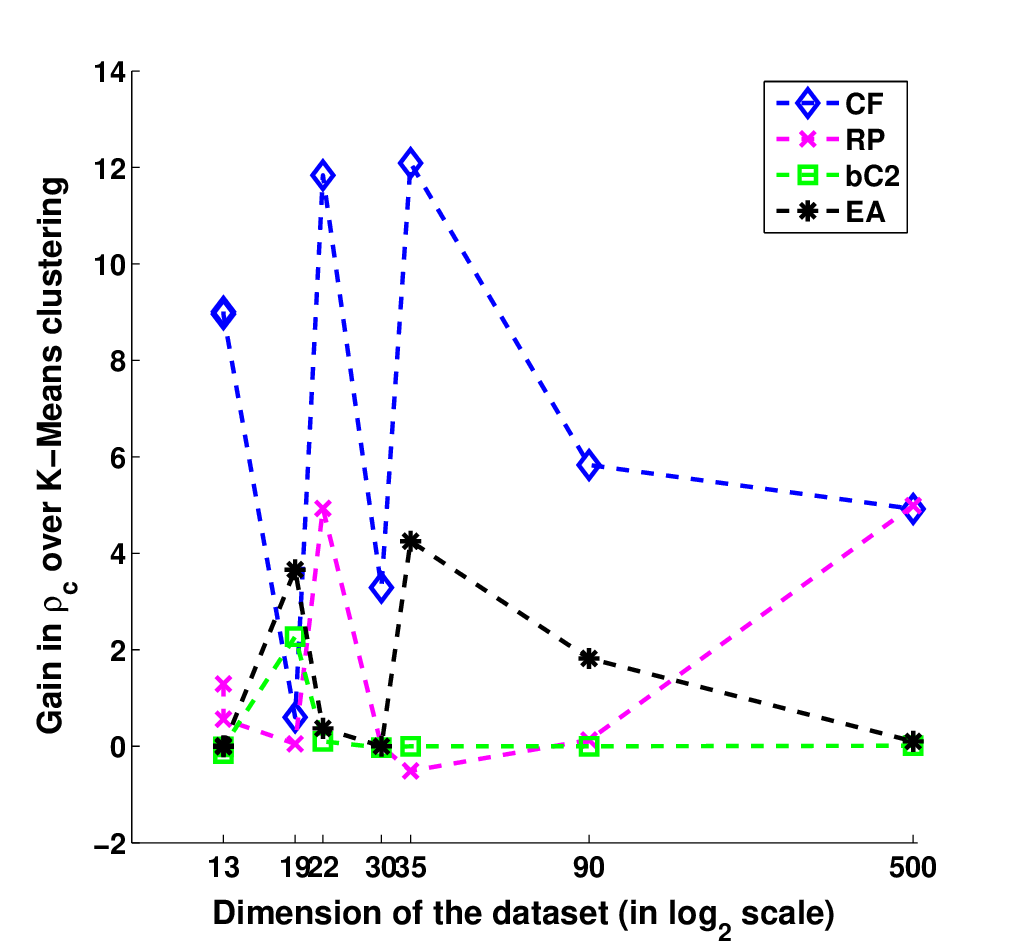} \vspace{-0.2in}\\
\end{center}
\caption{Performance gain of CF, RP, bC2 and EA over the baseline $K$-means
clustering algorithm according to $\rho_r$ and $\rho_c$, respectively. The
plot is arranged according to the data dimension of the eight UCI datasets.}
\label{figure:gainOverKmeans}
\end{figure}

Table~\ref{table:simulationResultsR} and
Table~\ref{table:simulationResultsC} show the values of $\rho_r$ and
$\rho_c$ (reported in percent throughout) achieved by different ensemble methods.
The ensemble size is $100$ and results averaged over 100
runs. We take $q=1$ for CF in producing these two tables. We see
that CF compares favorably to its competitors; it yields the largest
$\rho_r$ (or $\rho_c$) for six out of eight datasets and is very
close to the best on one of the other two datasets, and the
performance gain is substantial in some cases (i.e., in five cases).
This is also illustrated in Figure~\ref{figure:gainOverKmeans}.

We also explore the feature competition mechanism in the
initial round of CF (cf. Section~\ref{section:clusteringTree}).
According to Theorem~\ref{theorem:noiseResistent}, in cases where
there are many noise features or weak features, feature competition
will decrease the chance of obtaining a weak clustering instance,
hence a boost in the ensemble performance can be expected. In
Table~\ref{table:simulationResultsCompR} and
Table~\ref{table:simulationResultsCompC}, we report results for
varying $q$ in the feature competition step.
\vspace{0.1in}
\begin{table}[H]
\begin{center}
\begin{tabular}{r|cccc}
\hline
Dataset     &CF              &RP       &bC2          &EA        \\
\hline
Soybean     &\bf{84.43}           &71.83    &72.34        &76.59\\
SPECT       &\bf{68.02}           &61.11    &56.28        &56.55\\
ImgSeg      &48.24                &47.71    &49.91        &\bf{51.30}\\
Heart       &\bf{68.26}           &60.54    &59.10        &59.26\\
Wine        &\bf{79.19}           &70.79    &70.22        &70.22\\
WDBC        &\bf{88.70}           &85.41    &85.38        &85.41\\
Robot       &\bf{41.20}           &35.50    &35.37        &37.19\\
Madelon     &55.12           &\bf{55.19}    &50.20        &50.30\\
\hline
\end{tabular}
\end{center} \caption{$\rho_c$ for different
datasets and methods (CF calculated when $q=1$).}
\label{table:simulationResultsC}
\end{table}

\begin{figure}[H]
\centering
\begin{center}
\vspace{-0.2in}
\includegraphics*[scale=0.4,angle=-90,clip]{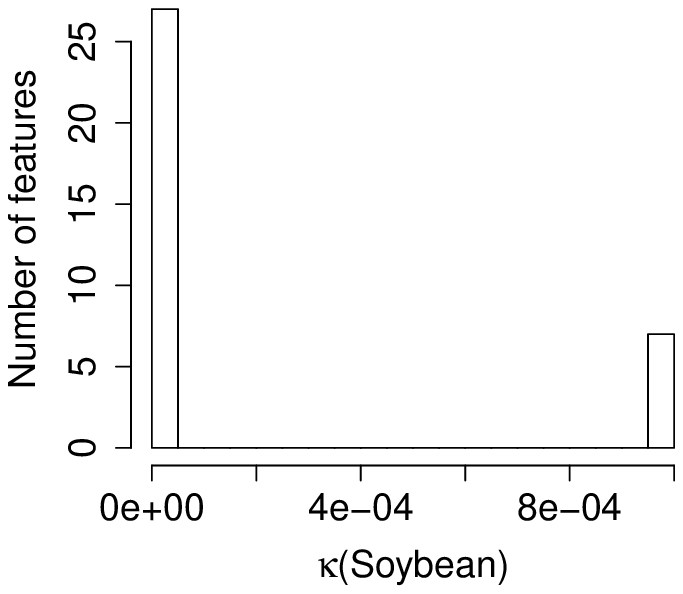}\quad\quad
\includegraphics*[scale=0.4,angle=-90,clip]{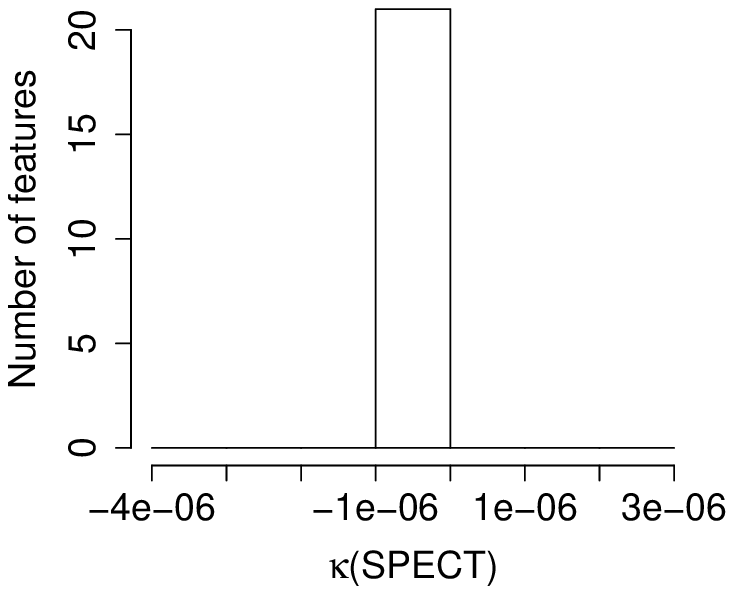} \vspace{-0.2in}\\
\includegraphics*[scale=0.4,angle=-90,clip]{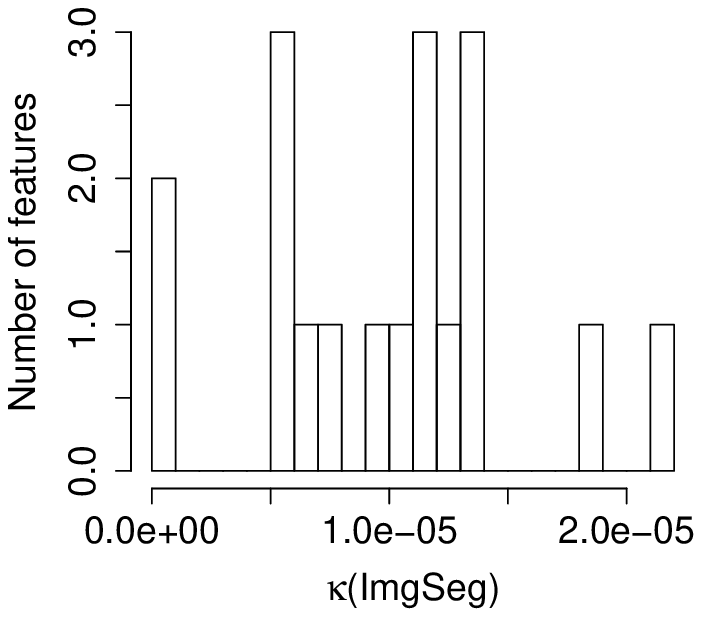}\quad\quad
\includegraphics*[scale=0.4,angle=-90,clip]{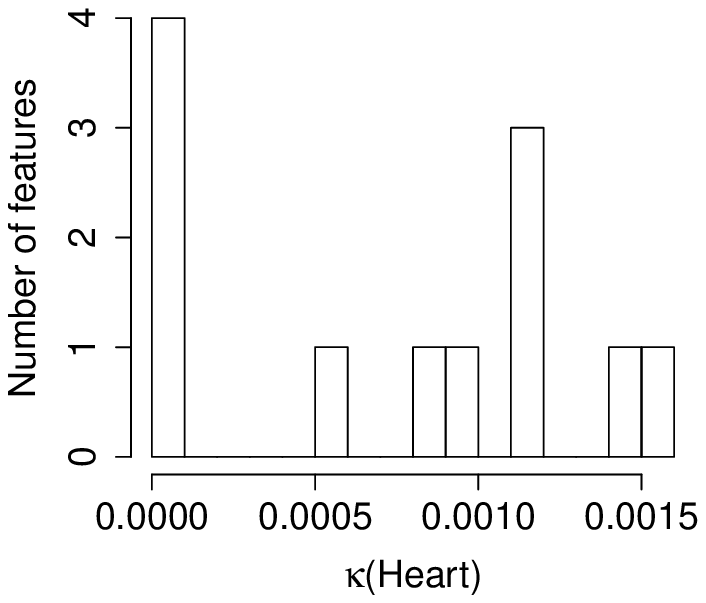} \vspace{-0.2in}\\
\includegraphics*[scale=0.4,angle=-90,clip]{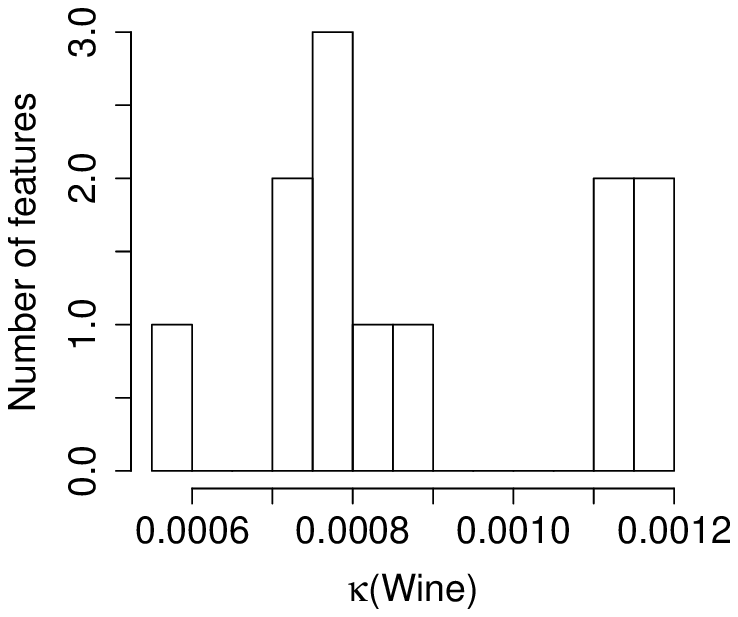}\quad\quad
\includegraphics*[scale=0.4,angle=-90,clip]{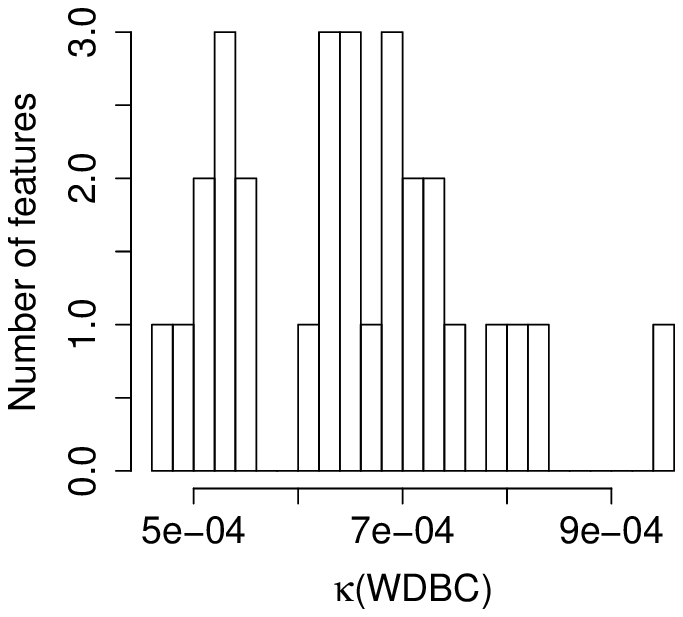}\vspace{-0.2in}\\
\includegraphics*[scale=0.4,angle=-90,clip]{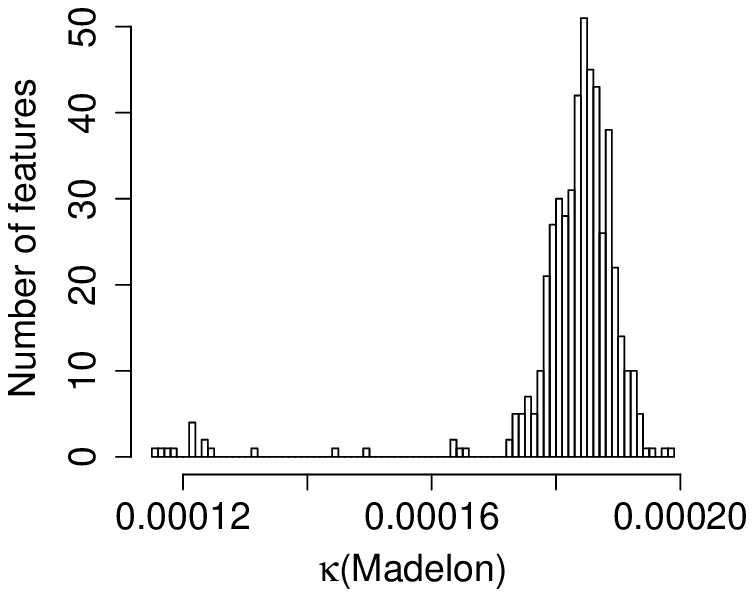}\quad\quad
\includegraphics*[scale=0.4,angle=-90,clip]{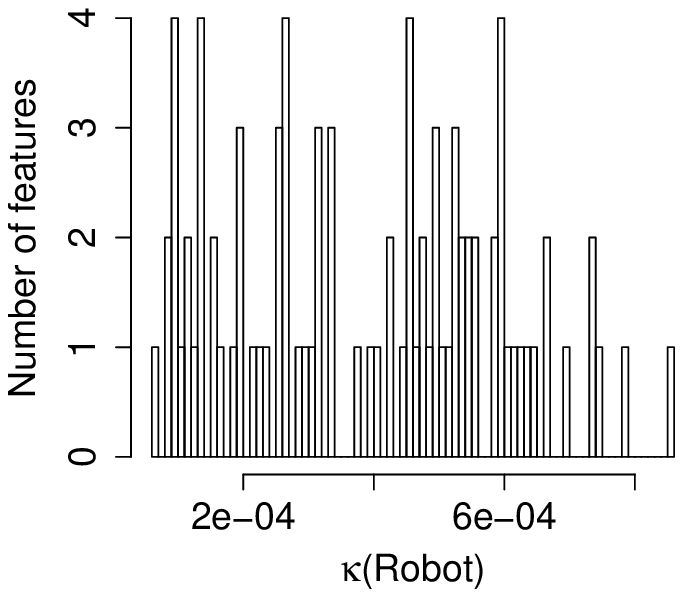}\vspace{-0.2in}\\
\end{center}
\caption{The feature profile plot for the eight UC Irvine datasets.}
\label{figure:fProfile}
\end{figure}

We define the \emph{feature profile plot} to be the histogram of the
strengths of each individual feature, where feature strength is
defined as the $\kappa$ value computed on the dataset using this
feature alone. (For categorical variables when the number of
categories on this variable is smaller than the number of clusters,
the strength of this feature is sampled at random from the set of
strengths of other features.) Figure~\ref{figure:fProfile} shows the
feature profile plot of the eight UC Irvine datasets used in our
experiment. A close inspection of results presented shows that this
plot can roughly guide us in choosing a ``good" $q$ for each individual
dataset. Thus a rule of thumb could be proposed: use large $q$ when there are many weak or
noise features and the difference in strength among features is big;
otherwise small $q$ or no feature competition at all. Alternatively,
one could use some cluster quality measure to choose $q$. For
example, we could use the $\kappa$ criterion as discussed in
Section~\ref{section:clusteringTree} or the Ncut value; we
leave an exploration of this possibility to future work.

\vspace{0.1in}
\begin{table}[H]
\begin{center}
\begin{tabular}{r|ccccccc}
\hline
$q$         &1           &2          &3        &5          &10         &15           &20\\
\hline
Soybean     &92.36       &92.32      &94.42   &93.89       &93.14      &94.54       &\bf{94.74}\\
SPECT       &56.78       &57.39      &57.24   &\bf{57.48}  &56.54      &55.62       &52.98\\
ImgSeg      &79.71       &77.62      &77.51   &81.17       &82.69      &\bf{83.10}  &82.37\\
Heart       &56.90       &60.08      &62.51   &63.56       &\bf{63.69} &\bf{63.69}  &\bf{63.69}\\
Wine        &\bf{79.70}  &74.02      &72.16   &71.87       &71.87      &71.87       &71.87\\
WDBC        &\bf{79.93}  &\bf{79.94} &79.54   &79.41       &78.90      &78.64       &78.50\\
Robot       &63.60       &63.86      &64.13   &64.75       &\bf{65.62} &65.58       &65.47\\
Madelon     &50.76       &\bf{50.94} &50.72   &50.68       &50.52      &50.40       &50.38\\
\hline
\end{tabular}
\caption{The $\rho_r$ achieved by CF for $q \in
\{1,2,3,5,10,15,20\}$. Results averaged over $100$ runs. Note the
first row is taken from Table~\ref{table:simulationResultsR}.}
\label{table:simulationResultsCompR}
\end{center}
\end{table}

\begin{table}[H]
\begin{center}
\begin{tabular}{r|ccccccc}
\hline
$q$         &1      &2       &3          &5          &10     &15    &20\\
\hline
Soybean    &84.43       &84.91        &89.85   &89.13       &88.40       &90.96      &\bf{91.91}\\
SPECT      &68.02       &\bf{68.90}   &68.70   &68.67       &66.99       &65.15      &60.87\\
ImgSeg     &48.24       &43.41        &41.12   &47.92       &49.77       &49.65      &\bf{52.79}\\
Heart      &68.26       &72.20        &74.93   &76.13       &\bf{76.30}  &\bf{76.30} &\bf{76.30}\\
Wine       &\bf{79.19}  &72.45        &70.52   &70.22       &70.22       &70.22      &70.22\\
WDBC       &\bf{88.70}  &\bf{88.71}   &88.45   &88.37       &88.03       &87.87      &87.75\\
Robot      &\bf{41.20}  &40.03       &39.57   &39.82       &38.40       &37.73       &37.68\\
Madelon    &55.12       &\bf{55.43}   &54.97   &54.92       &54.08       &53.57       &53.57\\
\hline
\end{tabular}
\caption{The $\rho_c$ achieved by CF for $q \in
\{1,2,3,5,10,15,20\}$. Results averaged over $100$ runs. Note the
first row is taken from Table~\ref{table:simulationResultsC}. }
\label{table:simulationResultsCompC}
\end{center}
\end{table}

Additionally, we also explore the effect of varying $n_b$, and
substantial performance gain is observed in some cases. For example,
setting $n_b=10$ boosts the performance of CF on ImgSeg to
$(\rho_c, \rho_r)=(62.34, 85.92)$, while $n_b=3$ on SPECT and WDBC
leads to improved $(\rho_c, \rho_r)$ at $(71.76, 59.45)$ and
$(90.00, 82.04)$, respectively. The intuition is that the initial
clustering by the base clustering algorithm may serve as a
pre-grouping of neighboring data points and hence achieves some
regularization with an improved clustering result. However, a
conclusive statement on this awaits extensive future work.

\subsubsection{Comparison to K-means clustering and spectral clustering}
\label{section:compKmeanSpec} We have demonstrated empirically that
CF compares very favorably to the three other clustering ensemble
algorithms (i.e., RP, bC2 and EA). One might be interested in how
much improvement CF achieves over the base clustering algorithm,
$K$-means clustering, and how CF compares to some of the ``best"

\noindent
clustering methods currently available, such as spectral clustering.
To explore this issue, we compared CF to $K$-means clustering and the
NJW spectral clustering algorithm (see \cite{NgJordan2002}) on the
eight UC Irvine datasets described in Table~\ref{table:datasets}. To
make the comparison with $K$-means clustering more robust, we run
$K$-means clustering under two different settings, denoted as
$K$-means-1 and $K$-means-2, where $(n_{it}, n_{rst})$ are taken as
$(200,20)$ and $(1000,100)$, respectively. For the NJW algorithm,
function {\it specc()} of the R project
package ``kernlab" (\cite{kernlab}) is used with the
Gaussian kernel and an automatic search of the local bandwidth
parameter. We report results under the two different clustering metrics, $\rho_c$ and
$\rho_r$, in Table~\ref{table:3comparisons}. It can be seen that CF improves
over $K$-means clustering on almost all datasets and the performance
gain is substantial in almost all cases.  Also, CF outperforms the
NJW algorithm on five out of the eight datasets.

\vspace{0.05in}
\begin{table}[h]
\begin{center}
\begin{tabular}{r|cccc}
\hline
Dataset     &CF             &NJW           &K-means-1         &K-means-2\\
\hline
Soybean     &\bf{92.36}     &83.72           &83.16             &83.16\\
            &\bf{84.43}     &76.60           &72.34             &72.34\\
SPECT       &\bf{56.78}     &53.77           &50.58             &50.58\\
            &\bf{68.02}     &64.04           &56.18             &56.18\\
ImgSeg      &79.71          &\bf{82.48}      &81.04             &80.97\\
            &48.24          &\bf{53.38}      &48.06             &47.21\\
Heart       &\bf{56.90}     &51.82           &51.53             &51.53\\
            &\bf{68.26}     &60.00           &59.25             &59.25\\
Wine        &\bf{79.70}     &71.91           &71.86             &71.86\\
            &\bf{79.19}     &70.78           &70.23             &70.22\\
WDBC        &79.93          &\bf{81.10}      &75.03             &75.03\\
            &88.70          &\bf{89.45}      &85.41             &85.41\\
Robot       &63.60          &\bf{69.70}      &39.76             &39.76\\
            &41.20          &\bf{42.68}      &35.37             &35.37\\
Madelon     &\bf{50.76}     &49.98           &49.98             &49.98\\
            &\bf{55.12}     &50.55           &50.20             &50.20\\
\hline
\end{tabular}
\end{center} \caption{Performance comparison between CF, spectral clustering,
and K-means clustering on the eight UC Irvine datasets. The
performance of CF is simply taken for $q=1$. The two numbers in each
entry indicate $\rho_r$ and $\rho_c$, respectively.}
\label{table:3comparisons}
\end{table}

\section{Conclusion}
\label{section:conclusion} We have proposed a new method for
ensemble-based clustering. Our experiments show that CF compares
favorably to existing clustering ensemble methods, including bC2,
evidence accumulation and RP.  The improvement of CF over the base
clustering algorithm (i.e., $K$-means clustering) is substantial,
and CF can boost the performance of $K$-means clustering to a level
that compares favorably to spectral clustering. We have provided
supporting theoretical analysis, showing that CF with $\kappa$ is
``noise-resistant" under a simplified model. We also obtain a
closed-form formula for the mis-clustering rate of spectral
clustering which yields new insights into the nature of spectral
clustering, in particular it underscores the importance of the
relative size of clusters to the performance of spectral clustering.





%




\section{Appendix}
In this appendix, Section~\ref{section:proofThm1} and
Section~\ref{section:proofThm2} are devoted to the proof of Theorem
1 and Theorem 2, respectively. Section~\ref{section:equivRules}
deals with the equivalence, in the population, of the optimal clustering
rule (as defined by equation (4) in Section 4.1 of the main text)
and $K$-means clustering. This is to prepare for the proof of
Theorem 1 and is of independent interest (e.g., it may help explain
why $K$-means clustering may be competitive on certain datasets in
practice). 
\subsection{Equivalence of $K$-means clustering and the optimal clustering rule
for mixture of spherical Gaussians} \label{section:equivRules}
\begin{figure}[h]
\centering
\begin{center}
\hspace{0cm}
\includegraphics*[scale=0.32,clip]{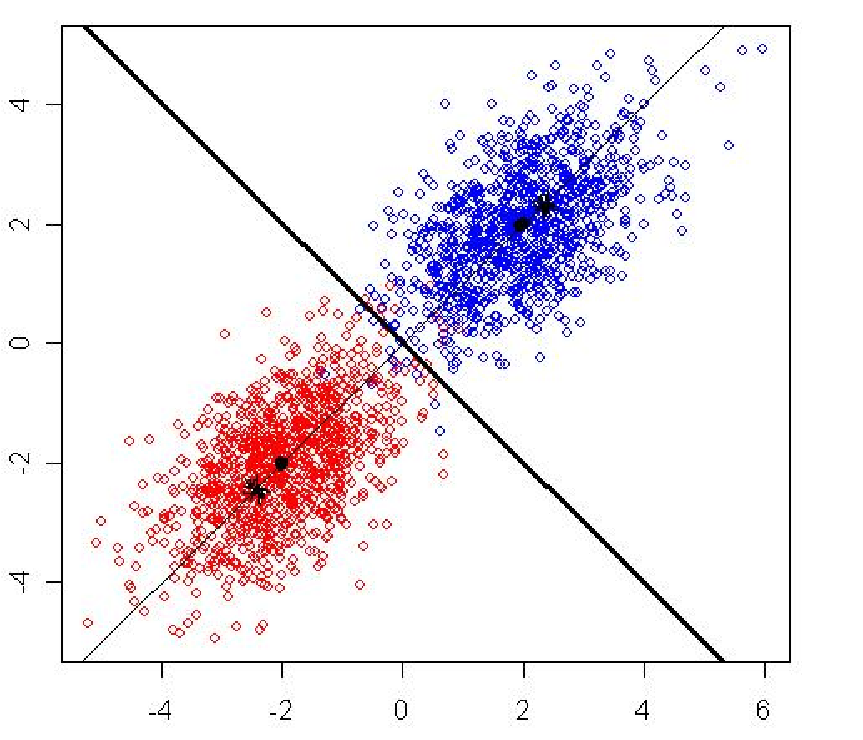}
\includegraphics*[scale=0.38,clip]{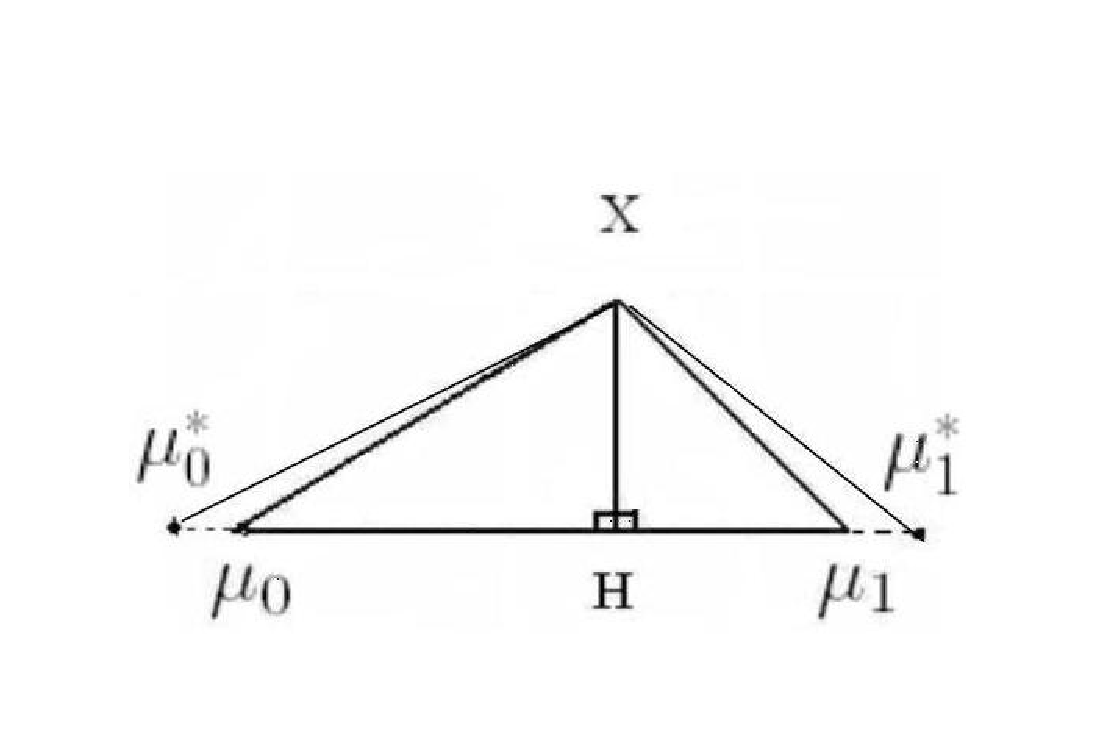}
\end{center}
\caption{\label{fig:scale} \it{The optimal rule $h^*$ and the
$K$-means rule. In  the left panel, the decision boundary (the thick
line) by $h^{*}$ and that by $K$-means completely overlap for a
2-component Gaussian mixture with $\Sigma=c\bm{I}$. The stars in the
figure indicate the population cluster centers by $K$-means. The
right panel illustrates the optimal rule $h^*$ and the decision rule
by $K$-means where $K$-means compares $||X-\mu_0^*||$ against
$||X-\mu_1^*||$ while $h^*$ compares $||H-\mu_0||$ against
$||H-\mu_1||$.}} \label{figure:optRule2}
\end{figure}

We first state and prove an elementary lemma for completeness.
\begin{lemma}
\label{lemma:equivalenceLemma} For the Gaussian mixture model defined by (2)
(Section~4.1) with $\Sigma=\bm{I}$ and $\pi=1/2$, in the population
the decision rule induced by $K$-means clustering (in the sense of
Pollard) is equivalent to the optimal rule $h^*$ as defined in (4)
(Section~4.1).
\end{lemma}
\noindent
\begin{proof}
The geometry underlying the proof is shown in
Figure~\ref{figure:optRule2}. Let $\mu_0, \Sigma_0$ and $\mu_1,
\Sigma_1$ be associated with the two mixture components in (2). By
shift-invariance and rotation-invariance (rotation is equivalent to
an orthogonal transformation which preserves clustering membership
for distance-based clustering), we can reduce to the $\mathbb{R}^1$
case such that $\bm{\mu}_0=(\mu_0[1],0,...0)=-\bm{\mu}_1$ with
$\Sigma_0=\Sigma_1=\bm{I}$. The rest of the argument follows from
geometry and the definition of $K$-means clustering, which assigns $X
\in \mathbb{R}^d$ to class $1$ if $||X-\mu_1^*||< ||X-\mu_0^*||$,
and the optimal rule $h^*$ which determines $X \in \mathbb{R}^d$ to
be in class 1 if
\begin{equation*}
\label{eq:opt} (\mu_1-\mu_0)^T(X-\frac{\mu_0+\mu_1}{2})>0,
\end{equation*}
or equivalently,
\begin{equation*}
||X-\mu_1|| < ||X-\mu_0||.
\end{equation*}
\end{proof}

\subsection{Proof of Theorem~1} \label{section:proofThm1}
\begin{proof}[Proof of Theorem~1]
Let $\mathcal{C}_1$ and $\mathcal{C}_0$ denote the two clusters
obtained by $K$-means clustering.  Let $G$ be the distribution
function of the underlying data.

$SS_W$ and $SS_B$ can be calculated as follows.
\begin{eqnarray*}
SS_W &=&\frac{1}{2}\int_{x \neq y \in \mathcal{C}_1} ||x-y||^2
dG(x)dG(y) + \frac{1}{2}\int_{x \neq y \in
\mathcal{C}_0} ||x-y||^2 dG(x)dG(y) \triangleq (\sigma^*_d)^2,\\
SS_B &=& \int_{x \in \mathcal{C}_1} \int_{y \in
\mathcal{C}_0}||x-y||^2 dG(y)dG(x) =
(\sigma^*_d)^2+\frac{1}{4}||\mu_0^*-\mu_1^*||^2.
\end{eqnarray*}
If we assume $\Sigma=\bm{I}_{p \times p}$ is always true during the
growth of the clustering vector (this holds if duplicated features
are excluded), then
\begin{equation}
\label{eq:kappaInverse}
\frac{1}{\kappa}=\frac{SS_B}{SS_W}=1+\frac{||\mu_0^*-\mu_1^*||^2}{4(\sigma^*_d)^2}.
\end{equation}
\noindent Without loss of generality, let
$\mathcal{I}=\{1,2,...,d-1\}$ and let the noise feature be the $d^{th}$
feature. By the equivalence, in the population, of $K$-means clustering
and the optimal clustering rule $h^*$ (Lemma~\ref{lemma:equivalenceLemma} in
Section~\ref{section:equivRules}) for a mixture of two spherical
Gaussians, $K$-means clustering assigns $x \in \mathbb{R}^d$ to
$\mathcal{C}_1$ if
\begin{equation*}
||x-\mu_1|| < ||x-\mu_0||,
\end{equation*}
which is equivalent to
\begin{equation}
\label{eq:kmeanNoiseRule} \sum_{i=1}^{d-1}
\left(x[i]-\mu_1[i]\right)^2 < \sum_{i=1}^{d-1}
\left(x[i]-\mu_0[i]\right)^2.
\end{equation}
This is true since $\mu_0[d]=\mu_1[d]=0$ by the assumption that the
$d^{th}$ feature is a noise feature. \eqref{eq:kmeanNoiseRule}
implies that the last coordinate of the population cluster centers
for $\mathcal{C}_1$ and $\mathcal{C}_2$ are the same, that is,
$\mu_1^{*}[d]=\mu_0^{*}[d]$. This is because, by definition,
$\mu_i^*[j]=\int_{x \in \mathcal{C}_i} x[j]dG(x)$ for $i=0,1$ and
$j=1,...,d$. Therefore adding a noise feature does not affect
$||\mu_0^*-\mu_1^*||^2$. However, the addition of a noise feature
would increase the value of $(\sigma^*_d)^2$, it follows that
$\kappa$ will be increased by adding a noise feature.
\end{proof}
\subsection{Proof of Theorem~2} \label{section:proofThm2}
\begin{proof}[Proof of Theorem 2.]
To simplify the presentation, some lemmas used here (Lemma~\ref{lem:bounds}
and Lemma~\ref{lem:semicircle-law}) are stated after this proof.

It can be shown that
\begin{eqnarray*}
D^{-1/2}\overline{P}D^{-1/2} & = &
\sum_{i=1}^{2}\lambda_{i}\mathbf{x}_{i}\mathbf{x}_{i}^{T},
\end{eqnarray*}
where $\lambda_{i}$ are the eigenvalues and $\mathbf{x}_{i}$
eigenvectors, $i=1,2$, such that for $\nu=o(1)$,
\begin{eqnarray*}
\lambda_{1} & = & 1+O_{p}(\nu^{2}+n^{-1}),\\
\lambda_{2} & = &
1-\gamma^{-1}(1+\gamma^{2})\nu+O_{p}(\nu^{2}+n^{-1})\end{eqnarray*}
 and
\[
(n_{1}\gamma^{-1}+n_{2})^{1/2}\mathbf{x}_{1}[i]=\left\{ \begin{array}{ll}
\gamma^{-1/2}+O_{p}(\nu+n^{-1/2}), & ~\mbox{if}~i\leq n_{1}\vspace{1pt}\\
1+O_{p}(\nu+n^{-1/2}), & ~\mbox{otherwise}.\end{array}\right.\]

\[
(n_{1}\gamma^{3}+n_{2})^{1/2}\mathbf{x}_{2}[i]=\left\{ \begin{array}{ll}
-\gamma^{3/2}+O_{p}(\nu+n^{-1/2}), & ~\mbox{if}~i\leq n_{1}\vspace{1pt}\\
1+O_{p}(\nu+n^{-1/2}), & ~\mbox{otherwise}.\end{array}\right.\]

By Lemma~\ref{lem:semicircle-law}, we have
$||\varepsilon||_{2}=O_{p}(\sqrt{n})$ and thus the $i^{th}$
eigenvalues of $D^{-1/2}PD^{-1/2}$ for $i\geq3$ are of order
$O_{p}(n^{-1/2})$. Note that, in the above, all residual terms are
uniformly bounded w.r.t. $n$ and $\nu$.

Let
\begin{eqnarray*}
\psi & = & \frac{1}{2\pi
i}\int_{\Gamma}(tI-D^{-1/2}PD^{-1/2})^{-1}dt,
\end{eqnarray*}
where $\Gamma$ is a Jordan curve enclosing only the first two
eigenvalues. Then, by (8) and (9) in the main text (see Section~4.2), 
$\psi\mathbf{x}_{2}$ is the second eigenvector of
$D^{-1/2}PD^{-1/2}$ and the mis-clustering rate is given by
\begin{equation*}
\mathcal{M} = \frac{1}{n} \left[ \sum_{i \leq n_1}
I((\psi\mathbf{x}_2)[i]<0) + \sum_{i > n_1}
I((\psi\mathbf{x}_2)[i]>0) \right].
\end{equation*}
Thus
\begin{equation*}
\mathbb{E}\mathcal{M} = \frac{1}{1+\gamma} \big[
\mathbb{P}((\psi\mathbf{x}_2)[i]>0) +
\gamma\mathbb{P}((\psi\mathbf{x}_2)[i]<0) \big].
\end{equation*}
By Lemma \ref{lem:bounds} and letting
$\tilde{\varepsilon}=D^{-1/2}\varepsilon D^{-1/2}$, we have
\begin{eqnarray*}
\psi\mathbf{x}_{2} & = & \frac{1}{2\pi i}\int_{\Gamma}\left(tI-D^{-1/2}\overline{P}D^{-1/2}-\tilde{\varepsilon}\right)^{-1}\mathbf{x}_{2}dt\\
 & = & \frac{1}{2\pi i}\int_{\Gamma}\left(I-(tI-D^{-1/2}\overline{P}D^{-1/2})^{-1}\tilde{\varepsilon}\right)^{-1}\left(tI-D^{-1/2}\overline{P}D^{-1/2}\right)^{-1}\mathbf{x}_{2}dt\\
 & = & \frac{1}{2\pi i}\int_{\Gamma}\left(I-(tI-D^{-1/2}\overline{P}D^{-1/2})^{-1}\tilde{\varepsilon}\right)^{-1}\mathbf{x}_{2}(t-\lambda_{2})^{-1}dt\\
 & = & \phi\mathbf{x}_{2}+O_{p}(n^{-2}),
 \end{eqnarray*}
where
\begin{eqnarray*}
\phi\mathbf{x}_{2} & = & \frac{1}{2\pi
i}\int_{\Gamma}\left[I+(tI-D^{-1/2}\overline{P}D^{-1/2})^{-1}\tilde{\varepsilon}\right]\mathbf{x}_{2}(t-\lambda_{2})^{-1}dt.
\end{eqnarray*}
It can be shown that, by the Cauchy Integral Theorem~\cite{RudinComplex1986} and Lemma \ref{lem:bounds},
\begin{eqnarray*}
\phi\mathbf{x}_{2} & = & \mathbf{x}_{2}+\frac{1}{2\pi i}\int_{\Gamma}\left(tI-D^{-1/2}\overline{P}D^{-1/2}\right)^{-1}\tilde{\varepsilon}\mathbf{x}_{2}(t-\lambda_{2})^{-1}dt\\
& = &
\mathbf{x}_{2}-\lambda_{2}^{-1}\tilde{\varepsilon}\mathbf{x}_{2}+O_{p}(n^{-2}).
\end{eqnarray*}
Let $\tilde{\varepsilon}_{i}$ be the $i^{th}$ column of
$\tilde{\varepsilon}$. By Slutsky's Theorem, 
one can verify that
\begin{eqnarray*}
\tilde{\varepsilon}_{1}^{T}\mathbf{x}_{2} & = &
\sigma(n_{1}n_{2})^{-1/2}\mathcal{N}\left(0,\frac{1+\gamma^{3}}{1+\gamma^{2}}\right)+O_{p}(n^{-2}),
\end{eqnarray*}
and
\begin{eqnarray*}
\tilde{\varepsilon}_{n}^{T}\mathbf{x}_{2} & =
&
n_{2}^{-1}\sigma\mathcal{N}\left(0,\frac{1+\gamma^{3}}{1+\gamma^{2}}\right)+O_{p}(n^{-2}).
\end{eqnarray*}
Thus
\begin{eqnarray*}
\mathbb{P}((\psi\mathbf{x}_2)[1]<0) &=& \mathbb{P}\left(\mathcal{N}(0,1)>(n_{1}n_{2})^{1/2}\sigma^{-1}\sqrt{\frac{1+\gamma^{2}}{1+\gamma^{3}}}\frac{\gamma^{3/2}}{\sqrt{n_{1}\gamma^{3}+n_{2}}}\right)(1+o(1))\\
& = &
\mathbb{P}\left(\mathcal{N}(0,1)>n^{1/2}\sigma^{-1}\sqrt{\frac{\gamma^{2}}{(1+\gamma)(1+\gamma^{3})}}\right)(1+o(1)),\end{eqnarray*}
and
\begin{eqnarray*}
\mathbb{P}((\psi\mathbf{x}_2)[1]>0) &=& \mathbb{P}\left(\mathcal{N}(0,1)>n_{2}\sigma^{-1}\sqrt{\frac{1+\gamma^{2}}{1+\gamma^{3}}}\frac{1}{\sqrt{n_{1}\gamma^{3}+n_{2}}}\right)(1+o(1))\\
& = &
\mathbb{P}\left(\mathcal{N}(0,1)>n^{1/2}\sigma^{-1}\sqrt{\frac{\gamma}{(1+\gamma)(1+\gamma^{3})}}\right)(1+o(1)).
\end{eqnarray*}
Hence
\begin{eqnarray*}
\lim_{n \rightarrow \infty}\frac{1}{n}\log (\mathbb{E}\mathcal{M})
&=& -\frac{\gamma^{2}}{2\sigma^2(1+\gamma)(1+\gamma^{3})},
\end{eqnarray*}
and the conclusion follows.
\end{proof}
\begin{lemma} \label{lem:bounds} Let $\overline{P},\mathbf{x}_{2}$,
$\lambda_{2}$, $\psi$, $\phi$ be defined as above. Then
\begin{eqnarray*}
\left(tI-D^{-1/2}\overline{P}D^{-1/2}\right)^{-1}\mathbf{x}_{2} & =
& (t-\lambda_{2})^{-1}\mathbf{x}_{2},
\end{eqnarray*}
and
\begin{eqnarray*}
||\psi\mathbf{x}_{2}-\phi\mathbf{x}_{2}||_{\infty} & = &
O_{p}(n^{-2}).
\end{eqnarray*}
\end{lemma}
\noindent The first part follows from a direct calculation and the
proof of the second relies on the semi-circle law. The technical
details are omitted.

\begin{lemma} \label{lem:semicircle-law} Let $\varepsilon=\{\varepsilon_{ij}\}_{i,j=1}^{n}$
be a symmetric random matrix with
$\varepsilon_{ij}\sim\mathcal{N}(0,1)$, independent for $1\leq i\leq
j\leq n.$ Then
\begin{eqnarray*}
||\varepsilon||_{2} & = & O_p(\sqrt{n}).
\end{eqnarray*}
\end{lemma}
\noindent The proof is based on the moment method (see
\cite{furedi.komlos.1981}) and the details are omitted. 


\bibliographystyle{plain}
\bibliography{cfBib}

\end{document}